\documentclass[aps,prd,twocolumn,showpacs]{revtex4}
\usepackage{graphicx}
\usepackage{dcolumn}
\usepackage{bm}
\begin{document}
\title{The effect of supersymmetry breaking in the Mass Varying
Neutrinos}

\author{Ryo Takahashi} 
\email{takahasi@muse.sc.niigata-u.ac.jp}
\affiliation{Graduate School of  Science and Technology,
 Niigata University,  950-2181 Niigata, Japan}
\author{Morimitsu Tanimoto}
\email{tanimoto@muse.sc.niigata-u.ac.jp}
\affiliation{Department of Physics,
 Niigata University,  950-2181 Niigata, Japan}

\date{\today}

\begin{abstract}
We discuss the effect of  the supersymmetry breaking 
on the Mass Varying Neutrinos(MaVaNs) scenario. Especially, the
effect mediated by the gravitational interaction between the hidden
sector and the dark energy sector is studied. A model including a
chiral superfield in the dark sector and the right handed neutrino
superfield is proposed. Evolutions of the neutrino mass and the
equation of state parameter are presented in the model. It is remarked
that only the mass of a sterile neutrino is variable in the case of
the vanishing mixing between the left-handed and a sterile neutrino on
cosmological time scale. The finite mixing makes the mass of the
left-handed neutrino variable.     
\end{abstract}


\maketitle

\section{Introduction}
In recent years, many cosmological observations have provided the
strong evidence that the Universe is flat and its energy density is
dominated by the dark energy component whose negative pressure causes
the cosmic expansion to accelerate \cite{Riess}$\sim$\cite{SDSS}. In
order to clarify the origin of the dark energy, one has tried to
understand the connection of the dark energy with particle physics.

In a dynamical model proposed by Fardon, Nelson and Weiner (MaVaNs),
relic neutrinos could form a negative pressure fluid and cause the
cosmic acceleration \cite{Weiner}. In this model, an unknown scalar
field which is called ``acceleron'' is introduced and neutrinos are
assumed to interact through a new scalar force. The acceleron sits at
the instantaneous minimum of its potential, and the cosmic expansion
only modulates this minimum through changes in the neutrino
density. Therefore, the neutrino mass is given by the acceleron, in
other words, it depends on its number density and changes with the
evolution of the Universe. The equation of state parameter $w$ and the 
dark energy density also evolve with the neutrino mass. Those
evolutions depend on a model of the scalar potential and the relation
between the acceleron and the neutrino mass strongly. Typical examples
of the potential have been discussed in ref. \cite{Peccei}.

The variable neutrino mass was considered at first in
ref. \cite{Yanagida}, and was discussed for neutrino clouds
\cite{McKellar}. Ref. \cite{Hung} considered coupling a sterile
neutrino to a slowly rolling scalar field which was responsible for
the dark energy. Ref. \cite{Wang} considered coupling of the dark
energy scalar, such as the quintessence to neutrinos and discuss its
impact on the neutrino mass limits from Baryogenesis. In the context
of the MaVaNs scenario, there have been a lot of works
\cite{Kaplan}$\sim$\cite{Zanzi}. The origin of the scalar potential
for the acceleron was not discussed in many literatures, however, that
is clear in the supersymmetric MaVaNs scenario \cite{our,SUSY}.

In this work, we present a model including the supersymmetry breaking
effect mediated by the gravity. Then we show evolutions of the
neutrino mass and the equation of state parameter in the model.

The paper is organized as follows: in Section II, we summarize the
supersymmetric MaVaNs scenario and present a model. Sec. III is
devoted to a discussion of the supersymmetry breaking effect mediated
by the gravity in the dark sector.  In Sec. IV, the summary is given.

\section{Supersymmetric MaVaNs}
In this section, we discuss the supersymmetric Mass Varying Neutrinos 
scenario and present a model.

The basic assumption of the MaVaNs with supersymmetry is to
introduce a chiral superfield $A$ in the dark sector, which is assumed 
to be a singlet under the gauge group of the standard model. It is
difficult to build a viable MaVaNs model without fine-tunings in some
parameters when one assumes one chiral superfield in the dark sector,
which couples to only the left-handed lepton doublet superfield
\cite{our}. Therefore, we assume that the superfield $A$ couples to
both the left-handed lepton doublet superfield $L$ and the
right-handed neutrino superfield $R$. 

In this framework, we suppose the superpotential
 \begin{eqnarray}
  W&=&\frac{\lambda _1}{6}A^3+\frac{M_A}{2}AA+m_DLA+M_DLR
      +\frac{\lambda _2}{2}ARR\nonumber\\
   &&+\frac{M_R}{2}RR,\label{W}
 \end{eqnarray}
where $\lambda _i(i=1,2)$ are coupling constant of $\mathcal{O}(1)$
and $M_A$, $M_D$, $M_R$ and $m_D$ are mass parameters. The scalar and
spinor component of $A$ are $(\phi ,\psi )$, and the scalar component
is assumed to be the acceleron which cause the present cosmic
acceleration. The spinor component is a sterile neutrino. The Helium-4
 abundancy gives the most accurate determination of a cosmological
 number of neutrinos and does not exclude a fourth thermalized
 neutrino at $T\sim \mbox{MeV}$ \cite{Strumia}. The third term of the
 right-hand side in Eq. (\ref{W}) is derived from the Yukawa coupling
 such as $yLAH$ with $y<H>=m_D$, where $H$ is the Higgs doublet. 

In the MaVaNs scenario, the dark energy is assumed to be the sum of
the neutrino energy density and the scalar potential for the acceleron:
 \begin{equation}
  \rho _{\mbox{{\scriptsize DE}}}=\rho _\nu +V(\phi ).
 \end{equation}
Since only the acceleron potential contributes to the dark energy, we
assume the vanishing vacuum expectation values of sleptons, and thus
the effective scalar potential is given as 
 \begin{equation}
  V(\phi )=\frac{\lambda _1^2}{4}|\phi |^4+M_A^2|\phi |^2
           +m_D^2|\phi |^2.
 \end{equation}
We can write down a lagrangian density from Eq. (\ref{W}),
 \begin{eqnarray}
  \mathcal{L}&=&\lambda _1\phi\psi\psi +M_A\psi\psi
                +m_D\bar{\nu}_L\psi+M_D\bar{\nu}_L\nu _R\nonumber\\
             &&+\lambda _2\phi\nu _R\nu _R+M_R\nu _R\nu _R+h.c. \ . 
  \label{lag}
 \end{eqnarray}
It is noticed that the lepton number conservation in the dark sector
is violated because this lagrangian includes both $M_A\psi\psi$ and
$m_D\bar{\nu}_L\psi$. After integrating out the right-handed neutrino,
the effective neutrino mass matrix is given by
 \begin{eqnarray}
  \mathcal{M}\simeq
   \left(
   \begin{array}{cc}
    -\frac{M_D^2}{M_R}+\frac{\lambda _2\phi M_D^2}{M_R^2} & m_D \\
    m_D & M_A+\lambda _1\phi
   \end{array}
  \right),
  \label{MM}
 \end{eqnarray}
in the basis of $(\bar{\nu}_L,\psi )$, where we assume $\lambda
_1\phi\ll M_D\ll M_R$. The first term of the $(1,1)$ element of this
matrix corresponds to the usual term given by the seesaw mechanism
\cite{seesaw,seesaw2,lowscale} in the absence of the acceleron. The
second term is derived from the coupling between the acceleron and the
right-handed neutrino but the magnitude of this term is negligible
small because of the suppression of $\mathcal{O}(1/M_R)$. Therefore,
we can rewrite the neutrino mass matrix as
 \begin{eqnarray}
  \mathcal{M}\simeq
   \left(
   \begin{array}{cc}
    c   & m_D \\
    m_D & M_A+\lambda _1\phi
   \end{array}
  \right),
 \end{eqnarray}
where $c\equiv -M_D^2/M_R$. It is remarked that only the mass of a
sterile neutrino is variable in the case of the vanishing mixing
($m_D=0$) between the left-handed and a sterile neutrino on
cosmological time scale. The finite mixing ($m_D\neq 0$) makes the
mass of the left-handed neutrino variable. We will consider these two
cases of $m_D=0$ and $m_D\neq 0$ later.

In the MaVaNs scenario, there are two constraints on the scalar
potential. The first one comes from observations of the Universe,
which is that the magnitude of the present dark energy density is
about $0.74\rho _c$, $\rho _c$ being the critical density. Thus, the
first constraint turns to
 \begin{eqnarray}
  V(\phi ^0)=0.74\rho _c-\rho _\nu ^0,
  \label{V}
 \end{eqnarray}
where ``$0$'' represents a value at the present epoch.

The second one is the stationary condition. In this scenario, the
neutrino mass is assumed to be a dynamical field which is a function
of the acceleron. Therefore, the dark energy density should be
stationary with respect to the variation of the neutrino mass:
 \begin{eqnarray}
  \frac{\partial\rho _{\mbox{{\scriptsize DE}}}}{\partial m_\nu}
  =\frac{\partial\rho _\nu}{\partial m_\nu}
   +\frac{\partial V(\phi (m_\nu))}{\partial m_\nu}=0.
  \label{stationary}
 \end{eqnarray}
If $\partial m_\nu/\partial\phi\neq 0$, this condition is equivalent
to the usual stationary condition stabilized by an ordinary scalar
field. Eq. (\ref{stationary}) is rewritten by using the cosmic
temperature $T$:
 \begin{eqnarray}
  \frac{\partial V(\phi )}{\partial m_\nu}
  =-T^3\frac{\partial F(\xi )}{\partial\xi},
  \label{stationary1}
 \end{eqnarray}
where $\xi\equiv m_\nu /T$, $\rho _\nu =T^4F(\xi )$ and
 \begin{eqnarray}
  F(\xi )\equiv\frac{1}{\pi ^2}\int _0^\infty
               \frac{dyy^2\sqrt{y^2+\xi ^2}}{e^y+1}.
 \end{eqnarray}
We can get the time evolution of the neutrino mass from
Eq. (\ref{stationary1}). Since the stationary condition should be
always satisfied in the evolution of the Universe, this one at the
present epoch is the second constraint on the scalar potential:
 \begin{eqnarray}
  \left.\frac{\partial V(\phi )}{\partial m_\nu}\right|
  _{m_\nu =m_\nu^0}
  =\left.-T^3\frac{\partial F(\xi )}{\partial\xi}\right|
   _{m_\nu =m_\nu^0,T=T_0}.
  \label{stationary2}
 \end{eqnarray} 
In addition to two constraints for the potential, we also have two
relations between the acceleron and the neutrino mass at the present
epoch:
 \begin{eqnarray}
  m_{\nu _L}^0&=&\frac{c+M_A+\lambda _1\phi ^0}{2}\nonumber\\
              &&+\frac{\sqrt{[c-(M_A+\lambda _1\phi ^0)]^2+4m_D^2}}
                      {2},\\
  m_\psi ^0&=&\frac{c+M_A+\lambda _1\phi ^0}{2}\nonumber\\
           &&-\frac{\sqrt{[c-(M_A+\lambda _1\phi ^0)]^2+4m_D^2}}{2}.
 \end{eqnarray}

Next, we will consider the dynamics of the acceleron field. In order
that the acceleron does not vary significantly on distance of
inter-neutrino spacing, the acceleron mass at the present epoch must
be less than $\mathcal{O}(10^{-4}\mbox{eV})$ \cite{Weiner}. Here and
below, we fix the present acceleron mass as
 \begin{eqnarray}
  m_\phi ^0=10^{-4}\mbox{ eV}.
  \label{amass}
 \end{eqnarray}
Once we adjust parameters which satisfy five equations (\ref{V}) and
(\ref{stationary2})$\sim$(\ref{amass}), we can have evolutions of the
neutrino mass by using the Eq. (\ref{stationary1}).  

The dark energy is characterized by the evolution of the equation of
state parameter $w$. The equation of state in this scenario is derived
from the energy conservation equation in the Robertson-Walker
background and the stationary condition Eq. (\ref{stationary1}):
 \begin{eqnarray}
  w+1=\frac{[4-h(\xi )]\rho _\nu}{3\rho _{\mbox{{\scriptsize DE}}}},
 \end{eqnarray}
where
 \begin{eqnarray}
  h(\xi )\equiv\frac{\xi\frac{\partial F(\xi )}{\partial\xi}}{F(\xi )}.
 \end{eqnarray}
It seems that $w$ in this scenario depend on the neutrino mass and the
cosmic temperature. This means that $w$ varies with the evolution of
the Universe unlike the cosmological constant.

In the last of this section, it is important to discuss the
hydrodynamic stability of the dark energy from MaVaNs. The speed of
sound squared in the neutrino-acceleron fluid is given by
 \begin{eqnarray}
  c_s^2=\frac{\dot{p}}
             {\dot{\rho}_{\mbox{{\scriptsize DE}}}}
       =\frac{\dot{w}\rho _{\mbox{{\scriptsize DE}}}
              +w\dot{\rho}_{\mbox{{\scriptsize DE}}}}
             {\dot{\rho}_{\mbox{{\scriptsize DE}}}},
 \end{eqnarray}
where $p$ is the pressure of the dark energy. Recently, it was argued
that when neutrinos are non-relativistic, this speed of sound squared
becomes negative in this scenario \cite{stability}. The emergence of
an imaginary speed of sound shows that the MaVaNs scenario with
non-relativistic neutrinos is unstable, and thus the fluid in this
scenario cannot acts as the dark energy. However, finite temperature
effects provide a positive contribution to the speed of sound squared
and avoid this instability \cite{Speed}. Then, a model should satisfy
the following condition,
 \begin{eqnarray}
  \frac{\partial m_\nu}{\partial z}
  \left(1-\frac{5aT^2}{3m_\nu ^2}\right)
  +\frac{25aT_0^2(z+1)}{3m_\nu}>0,
  \label{constraint}
 \end{eqnarray}
where $z$ is the redshift parameter, $z\equiv (T/T_0)-1$, and 
 \begin{eqnarray}
  a\equiv\frac{\int _0^\infty\frac{dyy^4}{e^y+1}}
              {2\int _0^\infty\frac{dyy^2}{e^y+1}}\simeq 6.47.
 \end{eqnarray}
Actually, some models satisfy this condition \cite{Honda,Spitzer}.

\section{The effect of supersymmetry breaking}
\begin{figure}[t]
\includegraphics[width=\linewidth]{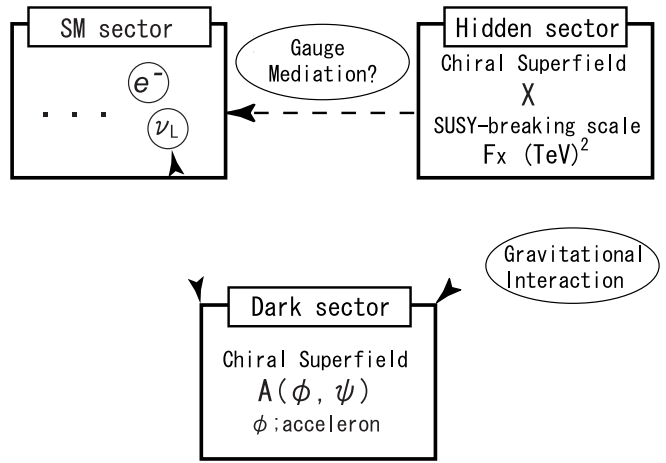}
\caption{The illustration of interactions among three sectors.
The dark sector couples to the left-handed neutrino through a new 
scalar force in the MaVaNs scenario. The dark sector is also assumed
to be related with the hidden sector only through the gravity.}
\label{fig:0}
\end{figure}

In order to consider the effect of supersymmetry breaking in the dark
sector, we assume a superfield $X$, which breaks supersymmetry, in the 
hidden sector, and the chiral superfield $A$ in the dark sector is
assumed to interact with the hidden sector only through the
gravity. This framework is shown graphically in
FIG. \ref{fig:0}. Once supersymmetry is broken at TeV scale, its
effect is transmitted to the dark sector through the following
operators:
 \begin{eqnarray}
  \int d^4\theta\frac{X^\dagger X}{M_{p\ell}^2}A^\dagger A,
  \hspace{3mm}
  \int d^4\theta\frac{X^\dagger +X}{M_{p\ell}}A^\dagger A,
 \end{eqnarray}
where $M_{p\ell}$ is the Planck mass. Then, the scale of the soft
terms $F_X(\mbox{TeV}^2)/M_{p\ell}\sim
\mathcal{O}(10^{-3}$-$10^{-2}\mbox{eV})$ is expected. Such a
framework was discussed in the ``acceleressence'' scenario
\cite{acceleressence}. Now, we consider only one superfield which
breaks supersymmetry for simplicity. If one extend the hidden sector,
one can consider a different mediation mechanism between the standard
model and the hidden sector from one between the dark and the hidden
sector. We will return to this point later.

In this framework, taking supersymmetry breaking effect into account,
the scalar potential is given by
 \begin{eqnarray}
  V(\phi )&=&\frac{\lambda _1^2}{4}|\phi |^4-\frac{\kappa}{3}(\phi ^3+h.c.)
             +M_A^2|\phi |^2+m_D^2|\phi |^2\nonumber\\
          &&-m^2|\phi |^2+V_0,
 \label{V1}
 \end{eqnarray}
where $\kappa$ and $m$ are supersymmetry breaking parameters, and $V_0$ is 
a constant determined by the condition that the cosmological constant
is vanishing at the true minimum of the acceleron potential. This
scalar potential is the same one presented in \cite{acceleressence}. We
consider two types of the neutrino mass matrix in this scalar
potential. They are the cases of the vanishing and the
finite mixing between the left-handed and a sterile neutrino.

\subsection{The Case of the Vanishing Mixing}
When the mixing between the left-handed and a sterile neutrino is
vanishing, $m_D=0$ in the neutrino mass matrix (\ref{MM}). Then we
have the mass of the left-handed and a sterile neutrino as
 \begin{eqnarray}
  m_{\nu _L} &=& c,\\
  m_\psi     &=& M_A+\lambda _1\phi. \label{sterilem}
\end{eqnarray}
In this case, we find that only the mass of a sterile neutrino is
variable on cosmological time scale due to the second term in
Eq. (\ref{sterilem}).

Let us adjust parameters which satisfy Eqs. (\ref{V}) and
(\ref{stationary2})$\sim$(\ref{amass}). In Eq. (\ref{V}), the scalar
potential Eq. (\ref{V1}) is used. Putting typical values for four
parameters by hand as follows:
 \begin{eqnarray}
  &&\lambda _1=1,\hspace{3mm}m_D=0,\nonumber\\
  &&m_{\nu _L}^0=2\times 10^{-2}\mbox{ eV},\hspace{3mm}
    m_\psi ^0=10^{-2}\mbox{ eV},
 \end{eqnarray}
we have
 \begin{eqnarray}
  &&\phi ^0\simeq -1.31\times 10^{-5}\mbox{ eV},\hspace{3mm}
    c=2\times 10^{-2}\mbox{ eV},\nonumber\\
  &&M_A\simeq 10^{-2}\mbox{ eV},\hspace{3mm}
    m\simeq 10^{-2}\mbox{ eV},\nonumber\\
  &&\kappa\simeq 4.34\times 10^{-3}\mbox{ eV}.\label{value}
 \end{eqnarray}
There is a tuning between $M_A$ and $m$ in order to satisfy the
constraint on the present accerelon mass of Eq. (\ref{amass}). 

\begin{figure}[t]
\includegraphics[width=\linewidth]{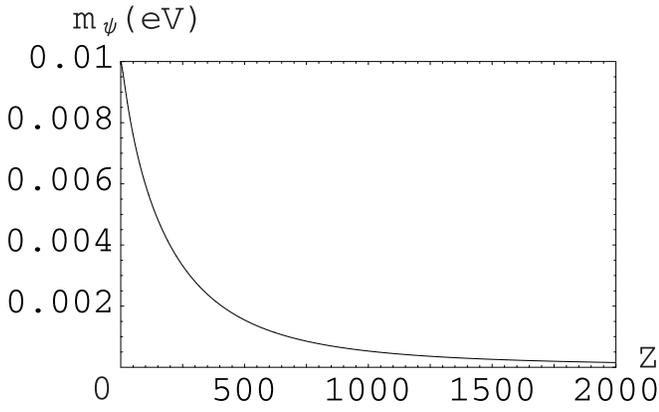}
\caption{Evolution of the mass of a sterile neutrino ($0\leq z\leq
2000$)}
\label{fig:2}
\end{figure}

\begin{figure}[t]
\includegraphics[width=\linewidth]{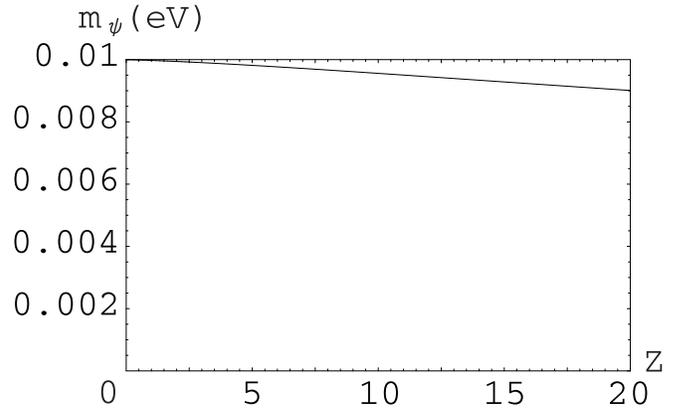}
\caption{Evolution of the mass of a sterile neutrino 
($0\leq z\leq 20$)}
\label{fig:3}
\end{figure}

\begin{figure}[t]
\includegraphics[width=\linewidth]{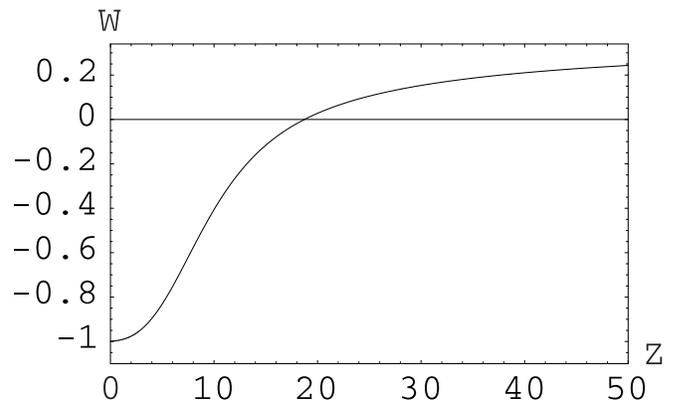}
\caption{Evolution of $w$ ($0\leq z\leq 50$)}
\label{fig:4}
\end{figure}

Now, we can calculate evolutions of the mass of a sterile neutrino and
the equation of state parameter $w$ by using these values. The
numerical results are shown in FIGs. \ref{fig:2}, \ref{fig:3} and
\ref{fig:4}. Especially, the behavior of the mass of a neutrino near
the present epoch is shown in FIG. \ref{fig:3}. We find that the mass 
of a sterile neutrino have varied slowly in this epoch. This means
that the first term of the left hand side in Eq. (\ref{constraint}),
which is a negative contribution to the speed of sound squared, is
tiny. We can also check the positive speed of sound squared in
a numerical calculation. Therefore, the neutrino-acceleron fluid is
hydrodynamically stable and acts as the dark energy.


\subsection{The Case of the Finite Mixing}

Next, we consider the case of the finite mixing between the
left-handed and a sterile neutrino ($m_D\neq 0$). In this case, the
left-handed and a sterile neutrino mass are given by
 \begin{eqnarray}
  m_{\nu _L}&=&\frac{c+M_A+\lambda _1\phi }{2}\nonumber\\
            &&+\frac{\sqrt{[c-(M_A+\lambda _1\phi)]^2+4m_D^2}}
                    {2},\\
  m_\psi&=&\frac{c+M_A+\lambda _1\phi }{2}\nonumber\\
        &&-\frac{\sqrt{[c-(M_A+\lambda _1\phi )]^2+4m_D^2}}{2}.
 \end{eqnarray}
We can expect that both the mass of the left-handed and a sterile
neutrino have varied on cosmological time scale due to the term of the 
acceleron dependence.
 
Taking typical values for four parameters as
 \begin{eqnarray}
  &&\lambda _1=1,\hspace{3mm}m_D=10^{-3}\mbox{ eV},\nonumber\\
  &&m_{\nu _L}^0=2\times 10^{-2}\mbox{ eV},\hspace{3mm}
    m_\psi ^0=10^{-2}\mbox{ eV},
 \end{eqnarray}
we have
 \begin{eqnarray}
  &&\phi ^0\simeq -1.31\times 10^{-5}\mbox{ eV},\hspace{3mm}
    c\simeq 1.99\times 10^{-2}\mbox{ eV},\nonumber\\
  &&M_A\simeq 1.01\times 10^{-2}\mbox{ eV},\hspace{3mm}
    m\simeq 1.02\times 10^{-2}\mbox{ eV},\nonumber\\
  &&\kappa\simeq 4.34\times 10^{-3}\mbox{ eV}.\label{value1}
 \end{eqnarray}
where we required that the mixing between the active and a sterile
neutrino is tiny. In our model, the small present value of the
 acceleron is needed to satisfy the constraints on the scalar
 potential in Eqs. (\ref{V}) and (\ref{stationary2}). 

\begin{figure}[h]
\includegraphics[width=\linewidth]{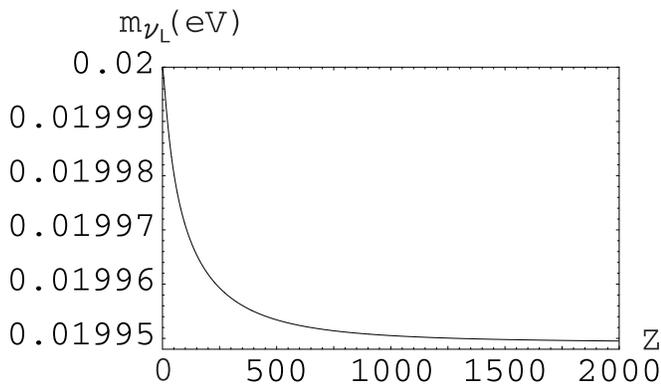}
\caption{Evolution of the mass of the left-handed neutrino
 ($0\leq z\leq 2000$)}
\label{fig:5}
\end{figure}
\begin{figure}[h]
\includegraphics[width=\linewidth]{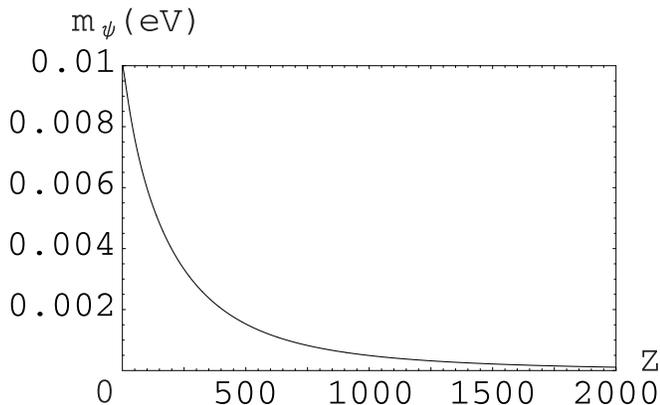}
\caption{Evolution of the mass of a sterile neutrino
 ($0\leq z\leq 2000$)}
\label{fig:6}
\end{figure}
\begin{figure}[h]
\includegraphics[width=\linewidth]{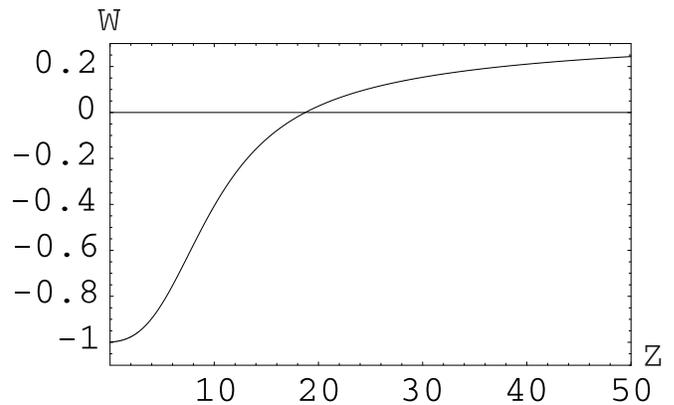}
\caption{Evolution of $w$ ($0\leq z\leq 50$)}
\label{fig:7}
\end{figure}

When we add following
 terms of the supersymmetry breaking effects to the scalar
 potential,
 \begin{eqnarray}
  \lambda _1m_{3/2}\phi^3,~~~M_Am_{3/2}|\phi |^2,
 \end{eqnarray}
a small gravitino mass ($m_{3/2}<\mathcal{O}(10\mbox{ eV})$) is
 favored. Such a small gravitino mass has been given in the gauge
 mediation model of Ref.\cite{Izawa}. Therefore, A mediation mechanism
 between the standard and the hidden sector which leads to a small
 gravitino mass is suitable for our framework. 

Values of parameters in (\ref{value1}) are almost same as the case of
the vanishing mixing (\ref{value}). However, the mass of the
left-handed neutrino is variable unlike the vanishing mixing case. The 
time evolution of the left-handed neutrino mass is shown in
FIG. \ref{fig:5}. The mixing does not affect the evolution of a
sterile neutrino mass and the equation of state parameter, which are
shown in FIGs. \ref{fig:6}, \ref{fig:7}. Since the variation in the
mass of the left-handed neutrino is not vanishing but extremely small,
the model can also avoid the instability of speed of sound. The value
of the sum of the left-handed and a sterile neutrino mass is within
the limit provided by analyses from WMAP three-years date
\cite{WMAP3,Fukugita}. 
 
Finally, we comment the smallness of the evolution of the neutrino
mass at the present epoch. In our model, the mass of the left-handed
and a sterile neutrino include the constant part. A variable part is a
function of the acceleron. In the present epoch, the constant part
dominates the neutrino mass because the present value of the acceleron
should be small. This smallness of the value of the acceleron is
required from the cosmological observation and the stationary
condition in Eqs. (\ref{V}) and (\ref{stationary2}). 
   
\section{Summary}
We presented a supersymmetric MaVaNs model including the effects of
the supersymmetry breaking mediated by the gravity. Evolutions of the
neutrino mass and the equation of state parameter have been calculated
in the model. Our model has a chiral superfield in the dark sector,
whose scalar component causes the present cosmic acceleration, and the
right-handed neutrino superfield. In our framework, supersymmetry is
broken in the hidden sector at TeV scale and the effect is assumed to
be transmitted to the  dark sector only through the gravity. Then, the 
scale of soft parameters are
$\mathcal{O}(10^{-3}$-$10^{-2})(\mbox{eV})$ is expected. 

We considered two types of model. One is the case of the vanishing
mixing between the left-handed and a sterile neutrino. Another one is
the finite mixing case. In the case of the vanishing mixing, only the
mass of a sterile neutrino had varied on cosmological time scale. In
the epoch of $0\leq z \leq 20$, the sterile neutrino mass had varied
slowly. This means that the speed of sound squared in the neutrino
acceleron fluid is positive, and thus this fluid can act as the dark
energy. In the finite mixing case, the mass of the left-handed
neutrino had also varied. However, the variation is extremely small
and the effect of the mixing does not almost affect the evolution of
the sterile neutrino mass and the equation of state. Therefore, this
model can also avoid the instability.

\end{document}